\documentstyle [12pt,epsfig] {article}               
\begin{document}

\begin{center}
{\bf BARYONS IN A CHIRAL CONSTITUENT QUARK MODEL}
\end{center}

\medskip

\begin{center}
L. Ya. Glozman
\end{center}

\begin{center}
{\it Institute for Theoretical Physics, 
        University of Graz,  Austria}
\end{center}

\begin{abstract}
In the low-energy regime
light and strange baryons should be considered
as systems of  constituent quarks with confining interaction and
a chiral interaction that is mediated by Goldstone bosons as well as
by vector  and scalar mesons.  The flavor-spin structure and sign of
the short-range part of the spin-spin force reduces the
$SU(6)_{FS}$ symmetry down to $SU(3)_F \times SU(2)_S$, induces hyperfine
splittings and provides correct ordering of the lowest states with positive
and negative parity. There is a cancellation of the tensor force from
pseudoscalar- and vector-exchanges in baryons. The spin-orbit interactions
from $\rho$-like and $\omega$-like exchanges also cancel each other in
baryons while they produce a big spin-orbit force in NN system. 
A unified description of 
light and strange baryon spectra calculated in a semirelativistic framework
is presented.  It is demonstrated that the same 
short-range part of spin-spin interaction between 
the constituent quarks 
 induces a strong short-range repulsion in $NN$ system when the
latter is treated as $6Q$ system.
Thus one can achieve a simultaneous understanding of a baryon structure
and baryon-baryon interaction in the low-energy regime. 
\end{abstract}

\section{Introduction}
Our aim in physics is not only to calculate some observable and get a
correct number but mainly to understand a physical picture responsible
for the given phenomenon. It  very often happens that a theory formulated
in terms of fundamental degrees of freedom cannot answer such a question
since it becomes overcomplicated at the related scale. Thus a main task in this
case is to select those degrees of freedom which are indeed essential.
For instance, the fundamental degrees of freedom in crystals are ions
in the lattice, electrons and the electromagnetic field. Nevertheless, in order
to understand electric conductivity, heat capacity, etc. we instead work
with  "heavy electrons" with dynamical mass, phonons and their interaction.
In this case a complicated electromagnetic interaction of the electrons with
the ions in the lattice is "hidden" in the dynamical mass of the electron
and the
interactions among ions in the lattice are eventually responsible  for the
 collective excitations of the lattice - phonons,
which are Goldstone bosons of the spontaneously broken translational
invariance in the lattice of ions.
As a result, the theory becomes rather
simple - only the electron and phonon degrees of freedom and their interactions
are essential for all the  properties of crystals mentioned above.

Quite a similar situation takes place in QCD. One hopes that sooner or later
one can solve the full nonquenched QCD on the lattice and get the
correct nucleon and
pion mass in terms of underlying degrees of freedom: current quarks and
gluon fields. However, QCD at the scale of 1 GeV becomes too complicated,
and hence it is rather difficult to say in this case what kind of
physics, inherent in
QCD, is relevant to the nucleon mass and its low-energy properties. In this
lecture I will try to answer this question. I will show that it is the
spontaneous breaking of chiral
symmetry  which is the most important QCD phenomenon
in this case, and that beyond the scale of spontaneous breaking of chiral
symmetry light and strange baryons can be viewed as systems of three
constituent quarks which interact by the exchange of Goldstone bosons
(pseudoscaler mesons), vector and scalar mesons (which could be
considered as a representation of a correlated Goldstone boson
exchange) and are subject to confinement.

\section{Spontaneous Breaking of Chiral Symmetry  and its Implications}

 At low temperatures and densities the  
$SU(3)_{\rm L} \times SU(3)_{\rm R}$ chiral symmetry of QCD Lagrangian
is spontaneously
broken down to $SU(3)_{\rm V}$ by the QCD vacuum (in the large $N_c$ limit
it would be $U(3)_{\rm L} \times U(3)_{\rm R} \rightarrow U(3)_{\rm V}$).
 A direct evidence for the spontaneously broken
chiral symmetry is a nonzero value of the quark condensates for the
light flavors
$<|\bar{q}q|> \approx -(240-250 {\rm
MeV})^3,$
which represent the order parameter. That this is indeed so, we know from
three  sources: current algebra,
QCD sum rules, and lattice gauge calculations.
There are two important generic consequences of the spontaneous breaking
of chiral symmetry (SBCS). The first one is an appearance of the octet
of pseudoscalar mesons of low mass, $\pi, {\rm K}, \eta$, which represent
the associated approximate Goldstone bosons (in the large $N_c$ limit the
flavor singlet state $\eta'$ should be added). The second one is that valence
(practically massless) quarks acquire a dynamical mass, which have been called 
historically  constituent quarks. Indeed, 
the nonzero value of the quark condensate 
itself implies at the formal level that there should be rather big
dynamical mass, which could be in general a moment-dependent quantity. 
Thus the constituent quarks should be considered as quasiparticles
whose dynamical mass comes from the nonperturbative 
gluon and quark-antiquark dressing.
The flavor-octet axial current conservation in the chiral limit tells that
the constituent quarks and Goldstone bosons should be coupled with the
strength $g=g_A M/f_\pi$ \cite{WEIN}, which is a quark analog of the famous 
Goldberger-Treiman relation. We cannot say at the moment for sure what is the
microscopical mechanism for SBCS in QCD. Any sufficiently strong
scalar interaction between quarks will induce the SBCS (e.g. the 
instanton - induced interaction contains the scalar part, or it can be
generated by monopole condensation, etc.).

All these general aspects of SBCS are well illustrated by the
Nambu and Jona-Lasinio model \cite{Nambu}, where the constituent mass
is generated by the scalar part of some nonperturbative local gluonic
interaction between current quarks while its pseudoscalar part
gives rise to a relativistic deeply-bound pseudoscalar
$Q\bar Q$ systems as Goldstone bosons.

Accordingly one arrives at the following interpretation of
light and strange baryons in the low-energy regime. The highly
nonperturbative gluodynamics gives rise to correlated 
quark-antiquark structures in the baryon sea (virtual mesons). 
At the same time
the current valence quarks get dressed by the quark condensates
and by the meson loops. The strongly-correlated quark-antiquark pairs
in the pseudoscalar channel manifest themselves 
by virtual pseudoscalar mesons, while the weakly-correlated
pairs in other channels - by vector, etc mesons. When one integrates
over the meson fields in the baryon wave function one arrives at the
simple QQQ Fock component with confined constituent quarks and with
residual interaction between them mediated by the corresponding meson
fields \cite{GLO2}. 

The complimentary description of the vector-meson fields as well as of
the scalar ones as arising from the correlated Goldstone bosons is
also possible, which does not contradict, however, to their interpretation
as weakly bound $Q\bar Q$ systems.

\section{The Goldstone Boson Exchange Interaction}

The
coupling of the constituent  quarks and the pseudoscalar Goldstone
bosons will (in the $SU(3)_{\rm F}$ symmetric approximation) have
the form $g/(2m)\bar\psi\gamma_\mu
\gamma_5\vec\lambda^{\rm F}
\cdot \psi \partial^\mu\vec\phi$ within the nonlinear realization of chiral
symmetry (it would be $ig\bar\psi\gamma_5\vec\lambda^{\rm F}
\cdot \vec\phi\psi$ within the linear $\sigma$-model chiral symmetry 
representation). A coupling of this
form, in a nonrelativistic reduction for the constituent quark spinors,
will -- to lowest order -- give rise the 
$\sim\vec\sigma \cdot \vec q \vec\lambda^{\rm F}$ structure of the 
meson-quark vertex, where $\vec q$ is meson momentum. Thus, the structure
of the  potential between quarks "$i$" and "$j$" in momentum representation is

\begin{equation}V(\vec q) \sim
\vec\sigma_i\cdot\vec q \sigma_j\cdot\vec q ~
\vec\lambda_i^{\rm F}\cdot\vec\lambda_j^{\rm F}
D(q^2) F^2(q^2)
,\label{2} \end{equation}

\noindent
where $D(q^2)$ is dressed Green function for chiral field which includes
both nonlinear terms of chiral Lagrangian and fermion loops, $F(q^2)$ is
meson-quark formfactor which takes into account the internal structure
of quasiparticles. At big distances ($\vec q \rightarrow 0$), one has 
$D(q^2) \rightarrow D_0({\vec q}^2)= -({\vec q}^2 + \mu^2)^{-1}
\not= \infty $ and $F(q^2) \rightarrow 1$.
 It then follows from
(\ref{2}) that $V(\vec q = 0) = 0$, which means that the volume integral
of the Goldstone boson exchange (GBE) interaction should vanish, 

\begin{equation}
\int d\vec r V(\vec r) = 0. 
\label{1} \end{equation}

\noindent
This sum rule is not valid in the chiral limit,
however, where $\mu =0$ and, hence, $D_0(\vec q =0)=\infty$.

The sum rule (\ref{1}) is trivial for the tensor component of the
pseudoscalar - exchange interaction since the tensor force component
automatically vanishes on averaging over the directions of $\vec r$.
But for the spin-spin component the sum rule  (\ref{1}) indicates that
there must be a strong short-range term. Indeed,
at big interquark separations the spin-spin component of the
pseudoscalar-exchange interaction is 
$V(r) \sim {e^{-\mu r}\over{r}}$, 
it then follows from the sum rule above
 that at short interquark separations the spin-spin
interaction should be opposite in sign as compared to the Yukawa tail and
very strong:

\begin{equation} H_\chi\sim -\sum_{i<j}V(\vec r_{ij})
\vec \lambda^{\rm F}_i \cdot \vec \lambda^{\rm F}_j\,
\vec
\sigma_i \cdot \vec \sigma_j.\label{1.1} \end{equation}
 
{\it It is this short-range part of the Goldstone boson exchange
(GBE) interaction between the constituent quarks that is of crucial importance
for baryons: it has a sign appropriate to reproduce the level splittings
and dominates over the Yukawa tail towards short distances.} In a
oversimplified consideration with a free Klein-Gordon Green function instead
of the dressed one in (\ref{2}) and with $F(q^2)=1$, one obtains the
following  spin-spin component of $Q-Q$ interaction:

\begin{equation}V(r)=
\frac{g^2}{4\pi}\frac{1}{3}\frac{1}{4m_im_j}
\vec\sigma_i\cdot\vec\sigma_j\vec\lambda_i^{\rm F}\cdot\vec\lambda_j^{\rm F}
\{\mu^2\frac{e^{-\mu r}}{ r}-4\pi\delta (\vec r)\}
.\label{3} \end{equation}

\noindent
In the chiral limit only the negative short-range part of the GBE interaction
survives.

\section{The Vector-Meson-Exchange Interaction}

As a representation of a multiple correlated Goldstone-boson-exchange
one can include  the exchange by a scalar flavor-singlet "$\sigma$"-meson
as well as by vector mesons.

The "$\sigma$"-meson exchange does not play a principal role in baryons
since it does not contain the spin-spin and tensor-force component. Its
attractive central potential as well as a weak spin-orbit force can be
effectively included into confining interaction in baryons. However,
these forces are of significant importance in NN system at medium range.

The coupling of constituent quarks with the octet vector meson field
$\vec v^{\mu}$ in the $SU(3)_F$ - symmetric approximation is

\begin{equation}L^v=
-g^v \bar\psi\gamma_\mu \vec\lambda^{\rm F} \cdot \psi \vec v^\mu
+ \frac {g^t}{2m} \bar\psi\sigma_{\mu \nu} \vec\lambda^{\rm F} \cdot 
\psi \partial^\nu \vec v^\mu,
\label{3.1} \end{equation}
 
\noindent
where $g^v$ and $g^t$ are the vector- and tensor coupling constants.
A coupling of this form - to lowest order - will give rise the spin-spin
, tensor, spin-orbit, and central interactions between the constituent
quarks.

It is very instructive to compare signs of the spin-spin and tensor
components of pseudoscalar- and vector-exchange interactions. The 
spin-spin and tensor components of the pseudoscalar-exchange interaction
arise from the $\vec \sigma \cdot \vec \nabla$ structure of the
pseudoscalar-meson -- constituent quark vertex as:

\begin{equation}
(\vec \sigma_i \cdot \vec \nabla)(\vec \sigma_j \cdot \vec \nabla)=
\frac {1}{3} (\vec \sigma_i \cdot \vec \sigma_j)\nabla^2 + \frac {1}{3}
\left[ 3(\vec \sigma_i \cdot \vec \nabla)(\vec \sigma_j \cdot \vec \nabla)
-  (\vec \sigma_i \cdot \vec \sigma_j)\nabla^2 \right].
\label{3.2} \end{equation}

The coupling of the vector-meson to constituent quark gives 
$\vec \sigma \times \vec \nabla$ structure of the vertex. Therefore the
spin-spin and tensor components of the vector-meson exchange interaction
arise as 

\begin{equation}
(\vec \sigma_i \times \vec \nabla)(\vec \sigma_j \times \vec \nabla)=
\frac {2}{3} (\vec \sigma_i \cdot \vec \sigma_j)\nabla^2 - \frac {1}{3}
\left[ 3(\vec \sigma_i \cdot \vec \nabla)(\vec \sigma_j \cdot \vec \nabla)
-   (\vec \sigma_i \cdot \vec \sigma_j)\nabla^2 \right].
\label{3.3} \end{equation}

Comparing (\ref{3.2}) with (\ref{3.3}) one observes that the spin-spin 
component of the vector-exchange has the same sign as the spin-spin
component of the pseudoscalar-exchange interaction and is "two times
stronger" while their tensor components have opposite signs.

The $\vec \sigma \times \vec \nabla$ structure  of the vector-meson --
constituent quark vertex also suggests that the spin-spin component
of the vector-meson-exchange interaction should satisfy the same
sum rule  (\ref{1}) as the pseudoscalar-exchange interaction. Then 
similar to pseudoscalar-exchange interaction there should be a 
short-range term in the vector-meson
exchange interaction of the form (\ref{1.1}). Again, if one neglects
the spatial structure of the meson-quark vertex and uses a free
Green function for the vector-meson field, this short range term is 
described by a $\delta$ - function piece similar to equation (\ref{3}).

Summarizing, both vector- and pseudoscalar-exchange interactions produce
the short-range spin-flavor force (\ref{1.1}) while their tensor forces
largely cancel each other in baryons. This observation is crucial
for the present model.\\

The spin-orbit component associated with the vector-meson exchange
interaction is big and empirically very important in NN system. So
the question is why this spin-orbit force is big in NN system and
becomes inessential in baryons, where the spin-orbit splittings are
generally very small (see e.g. at $N(1535) - N(1520)$,
 $N(1650) - N(1700) - N(1675)$, ... LS multiplets). The reason for
this remarkable phenomenon is an explicit flavor dependence of the
vector-meson-exchange LS force.

Consider for simplicity the $SU(2)_T$ case, i.e. an exchange by 
$\rho$- and $\omega$-mesons between U and D quarks. The $\rho$-meson
is isovector and the $\rho$-meson exchange potential contains
the factor $\vec \tau_i \cdot \vec \tau_j$. The $\omega$-meson is
isoscalar and the $\omega$-exchange interaction does not contain
the isospin-dependent factor. In $^3P_J$ NN partial wave the isospin
of the two-nucleon system is $T=1$ and the isospin matrix element for
the $\rho$-meson exchange potential is 
$<T=1|\vec \tau_i \cdot \vec \tau_j|T=1> = 1$. Thus both $\omega$-
and $\rho$-meson exchange spin-orbit forces contribute with the same sign
in NN system. Numerically the contribution from $\omega$-exchange is 
2.5-3 times bigger than from the $\rho$-meson exchange spin-orbit force.

Now let us consider $^3P_J$ state of two light quarks in baryon. In the
present case the isospin of the quark pair is $T=0$, due to the presence
of the color part of wave function. Then for the $\rho$-meson exchange 
one obtains $<T=0|\vec \tau_i \cdot \vec \tau_j|T=0> = -3$. Thus the
$\rho$-meson exchange spin-orbit force obtains opposite sign and becomes
strongly enhanced in baryons. As a result one observes a strong
cancellation of the $\rho$- and $\omega$-meson exchange spin-orbit forces 
in baryons and the net weak spin-orbit interaction does not induce
 appreciable splittings in baryons.

Numerical details about both the tensor- and spin-orbit force cancellation
in baryons can be found in ref. \cite{GPVWVECTOR}.

\section{The Flavor-Spin Hyperfine Interaction and the Structure
of the Baryon Spectrum}

 Summarizing previous sections one concludes that the pseudoscalar-
and vector-meson exchange interactions produce strong 
flavor-spin interaction (\ref{1.1}) at short range while the net
tensor and spin-orbit forces are rather weak. That the net spin-orbit
and tensor interactions between constituent quarks {\it in baryons}
should be weak also follows from the typically small splittings
in LS-multiplets, which are of the order 10-30 MeV. These small splittings
should be compared with the hyperfine splittings produced by spin-spin
force, which are of the order of $\Delta - N$ splitting. Thus, indeed,
in baryons it is the spin-spin interaction  (\ref{1.1}) between constituent
quarks is of crucial importance.

Consider first, for the purposes of illustration, a schematic model
which neglects the radial dependence
of the potential function $V(r)$ in (\ref{1.1}), and assume a harmonic
confinement among quarks as well as $m_{\rm u}=m_{\rm d}=m_{\rm s}$.
In this model

\begin{equation}H_\chi = -\sum_{i<j}C_\chi~
\vec \lambda^{\rm F}_i \cdot \vec \lambda^{\rm F}_j\,
\vec
\sigma_i \cdot \vec \sigma_j.\label{4} \end{equation}

\noindent
Note, that contrary to the color-magnetic interaction from perturbative
one-gluon exchange, the chiral interaction is explicitly flavor-dependent.
It is this circumstance which allows to solve the long-standing problem
of ordering of the lowest positive-negative parity states.

If the only interaction between the
quarks were the flavor- and spin-independent harmonic confining
interaction, the baryon spectrum would be organized in multiplets
of the symmetry group $SU(6)_{\rm FS} \times U(6)_{\rm conf}$. In this case
the baryon masses would be determined solely by the orbital structure,
and the spectrum would be organized in an {\it alternative sequence
of positive and negative parity states,} i.e. in this case the spectrum
would be: ground state of positive parity 
($N=0$ shell, $N$ is the number of harmonic 
oscillator excitations in a 3-quark state), first excited band of
negative parity ($N=1$), second excited band of positive parity ($N=2$), etc.

The Hamiltonian (\ref{4}), within a first order perturbation theory,
 reduces the $SU(6)_{\rm FS} \times U(6)_{\rm conf}$ symmetry down to
 $SU(3)_{\rm F}\times SU(2)_{\rm S}\times U(6)_{\rm conf}$, which automatically
implies a splitting between the octet and decuplet baryons (e.g. the $\Delta$
resonance becomes heavier than nucleon).

Let us now see how the pure confinement spectrum above becomes modified when
the GBE Hamiltonian (\ref{4}) is switched on.
For the octet states ${\rm N}$, $\Lambda$, $\Sigma$,
$\Xi$ ($N=0$ shell) as well as for their first
radial excitations of positive parity
 ${\rm N}(1440)$, $\Lambda(1600)$, $\Sigma(1660)$,
$\Xi(?)$ ($N=2$ shell) 
 the expectation value of the
Hamiltonian (\ref{4})
 is $-14C_\chi$. For the decuplet states
$\Delta$, $\Sigma(1385)$, $\Xi(1530)$, $\Omega$ ($N=0$ shell)
the corresponding matrix element is
$-4C_\chi$. In the  negative parity excitations
($N=1$ shell) in the ${\rm N}$, $\Lambda$ and $\Sigma$ spectra 
(${\rm N}(1535)$ - ${\rm N}(1520)$, $\Lambda(1670)$ - $\Lambda(1690)$
and $\Sigma(1750)$ - $\Sigma(?)$)
the contribution of the interaction (\ref{4})  is $-2C_\chi$. 
The first negative
parity excitation in the $\Lambda$ spectrum ($N=1$ shell)
$\Lambda(1405)$ - $\Lambda(1520)$ is flavor singlet 
and, in this case, the corresponding matrix element is $-8C_\chi$. The latter
state is unique and is absent in other spectra due to its flavor-singlet
nature.

These  matrix elements alone suffice to prove that
the ordering of the lowest positive and negative parity states
in the baryon spectrum will be correctly predicted by
the chiral boson exchange interaction (\ref{4}).
The constant $C_\chi$ may be determined from the
N$-\Delta$ splitting to be 29.3 MeV.
The oscillator
parameter $\hbar\omega$, which characterizes the
effective confining interaction,
may be determined as  one half of the mass differences between the
first excited
$\frac{1}{2}^+$ states and the ground states of the baryons,
which have the same flavor-spin, flavor and spin symmetries
(e.g. ${\rm N}(1440)$ - ${\rm N}$, $\Lambda(1600)$ - $\Lambda$, $\Sigma(1660)$
- $\Sigma$),
to be
$\hbar\omega \simeq 250$ MeV. Thus the two free parameters of this simple model
are fixed and we can  make now predictions.

In the ${\rm N}$, $\Lambda$  and $\Sigma$ sectors the mass
difference between the lowest
excited ${1\over 2}^+$ states (${\rm N}(1440)$, $\Lambda(1600)$, 
and $\Sigma(1660)$)
and ${1\over 2}^--{3\over 2}^-$ negative parity pairs
 (${\rm N}(1535)$ - ${\rm N}(1520)$, $\Lambda(1670)$ - $\Lambda(1690)$,
 and $\Sigma(1750)$ - $\Sigma(?)$, respectively) will then
be
\begin{equation}{\rm N},\Lambda,\Sigma:
\quad m({1\over 2}^+)-m({1\over 2}^--{3\over
2}^-)=250\, {\rm
MeV}-C_\chi(14-2)=-102\, {\rm MeV},\end{equation}
whereas for the lowest states in the $\Lambda$ system ($\Lambda(1600)$,
$\Lambda(1405)$ - $\Lambda(1520)$) it should be

\begin{equation}\Lambda:\quad m({1\over 2}^+)-m({1\over 2}^--{3\over
2}^-)=250\, {\rm
MeV}-C_\chi(14-8)=74\, {\rm MeV}.  \end{equation}

This simple example shows how the chiral interaction 
provides different ordering of the lowest positive and negative parity excited
states in the spectra of the nucleon and
the $\Lambda$-hyperon. This is a direct
consequence of the symmetry properties of the boson-exchange interaction
\cite{GLO2}.
Namely, completely symmetric FS state in the ${\rm N}(1440)$,
$\Lambda(1600)$ and
$\Sigma(1660)$ positive parity resonances from the $N=2$ band feels a
much stronger
attractive interaction than the mixed symmetry state  in the
${\rm N}(1535)$ - ${\rm N}(1520)$, $\Lambda(1670)$ - $\Lambda(1690)$
and $\Sigma(1750)$ -$\Sigma(?)$ resonances of negative parity ($N=1$ shell).
Consequently the masses of the
positive parity states ${\rm N}(1440)$, $\Lambda(1600)$  and
$\Sigma(1660)$ are shifted
down relative to the other ones, which explains the reversal of
the otherwise expected "normal ordering".
The situation is different for $\Lambda(1405)$ - $\Lambda(1520)$
and
$\Lambda(1600)$, as the flavor state of  $\Lambda(1405)$ - $\Lambda(1520)$ is
totally antisymmetric. Because of this the
$\Lambda(1405)$ - $\Lambda(1520)$ gains an
attractive energy, which is
comparable to that of the $\Lambda(1600)$, and thus the ordering
suggested by the confining oscillator interaction is maintained.

Note that the problem of the relative position of positive-negative
parity states cannot be solved with other types of hyperfine interactions
between constituent quarks (the colour-magnetic and instanton-induced ones).

\section{Semirelativistic Chiral Constituent Quark Model}

In the semirelativistic chiral constituent quark model \cite{GPVW,GPVWVECTOR}
the dynamical part of the Hamiltonian consists of linear pairwise
confining interaction with the string tension fixed to the known
value 1 GeV/fm from Regge slopes (which also follows from
the heavy quarkonium spectroscopy and lattice calculations), 
and the chiral interaction, mediated by pseudoscalar,
scalar and vector-meson exchanges.  
Both the flavor-octet  and flavor-singlet
 pseudoscalar- and vector-meson exchanges are taken into account. 
The coupling constants of constituent quarks with mesons
are fixed from the empirically known 
$\pi N$, $\rho N$, and $\omega N$ coupling constants. The "sigma-meson" --
constituent quark
coupling constant is taken to be equal the pion -- constituent quark
coupling constant, as constrained by chiral symmetry.
For the constituent quark masses one takes typical values 
$M_{u,d} =340$ MeV, $M_s = 500$ Mev
and do not fit them. The short-range behaviour of the interaction is
determined by the cut-off parameters $\Lambda$ 
in the constituent quark -- meson
form-factors, which are taken in monopole form. In order to avoid a
proliferation of free parameters, by assuming independent values of 
$\Lambda$ for each meson, we adopt the linear scaling prescription
$\Lambda = \Lambda_0 + \kappa \mu$, where $\mu$ is meson mass, for
pseudoscalar- and vector mesons.

The kinetic-energy operator is taken in relativistic form,
$H_0 = \sum_{i=1}^3 \sqrt(p_i^2 + m_i^2)$. The semirelativistic 
three-quark Hamiltonian was solved along the stochastical variational
method \cite{VARGA} in momentum space. For the whole Q-Q potential the
model involves a total of 4 free parameters whose numerical values are
determined from the fit to all 35 confirmed low-lying states.

In fig. 1 we present the ground states as well as low-lying excited states
in $N$, $\Delta$, $\Lambda$, $\Sigma$, $\Xi$, and $\Omega$ spectra.
From the results of fig. 1 it becomes evident that within the chiral
constituent quark model a unified description of both nonstrange and
strange baryon spectra is achieved in good agreement with phenomenology.

It is instructive to learn how the spin-spin 
interaction  (\ref{1.1}) affects the energy levels 
when it is switched on and its strength is gradually increased (Fig. 2).
Starting out from the case with confinement only, one observes that
the degeneracy of states is removed and the inversion of ordering
of positive and negative parity states 
is achieved in the $N$ spectrum, as well as for some states in the
$\Lambda$ spectrum, while the ordering of the lowest positive-negative
parity states is opposite in $N$ and $\Lambda$ spectra.

\begin{figure}
\begin{center}
\epsfig{file=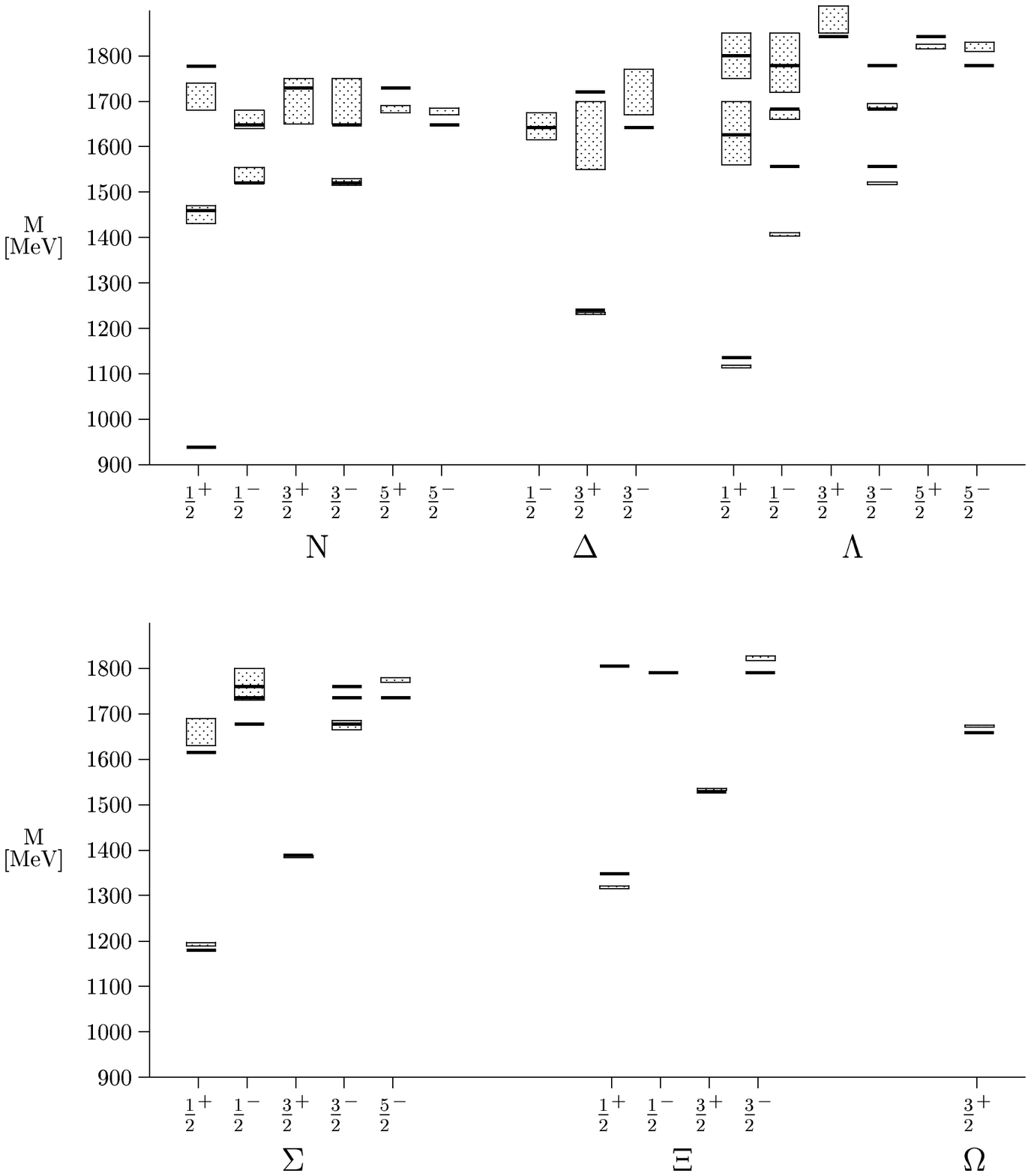}
\caption{Energy levels of the lowest light and strange baryon states 
(below 1850 MeV) with total
angular momentum and parity $J^P$. The shadowed boxes represent the experimental
values with their uncertainties.}
\end{center}
\end{figure}
\begin{figure}
\epsfig{file=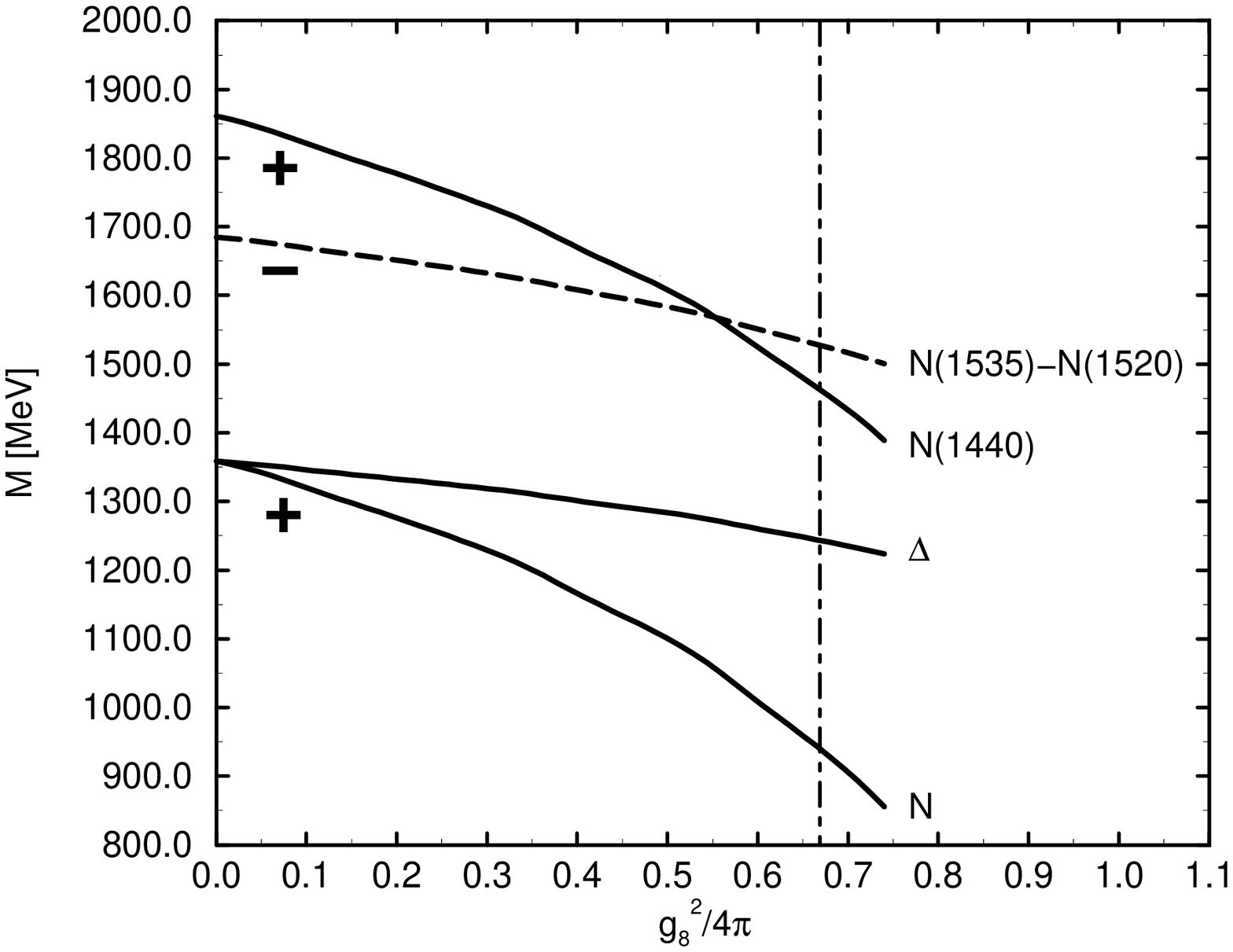,width=8cm}\hfill
\epsfig{file=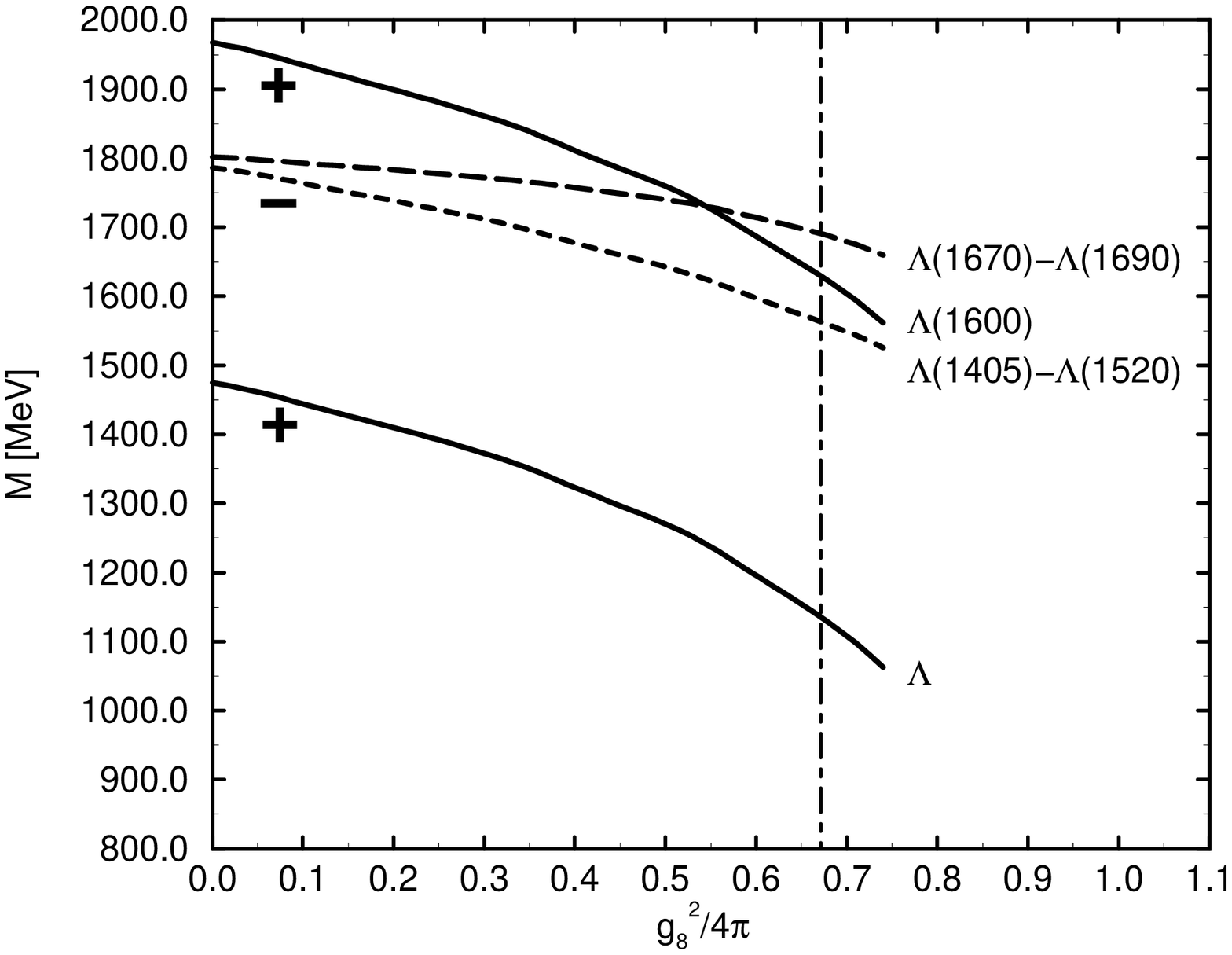,width=8cm}
\caption{Level shifts of some lowest baryons as a function of the 
strength of the chiral interaction. 
Solid and dashed lines correspond to positive- 
and negative-parity states, respectively.}
\end{figure}
\begin{figure}
\begin{center}
\epsfig{file=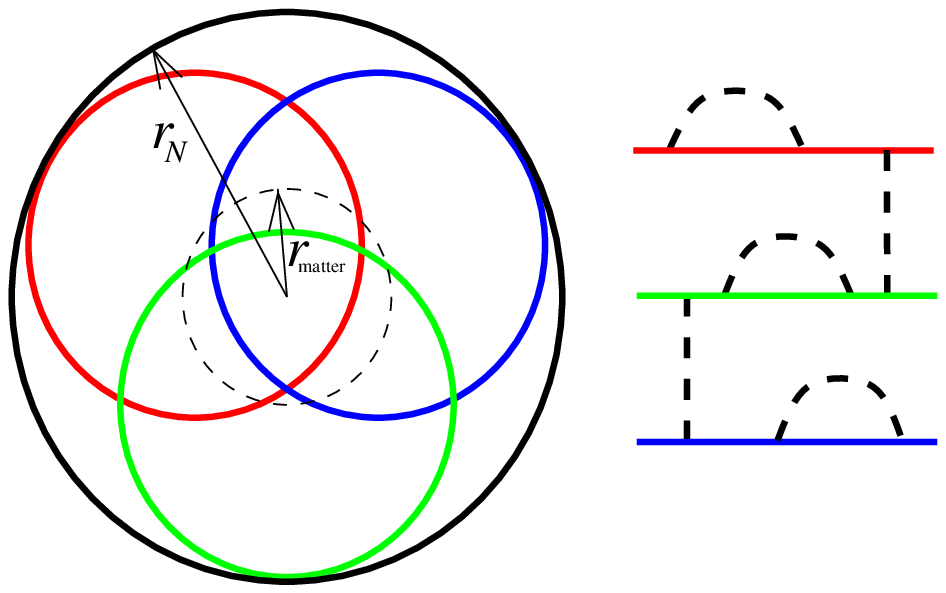}
\caption{Nucleon as it is seen in the low-energy and low-resolution regime.} 
\end{center}
\end{figure}

It is clear that the Fock components $QQQ\pi, QQQK, ...$
(including meson continuum) cannot be completely integrated out in favour
of the meson-exchange $Q-Q$ potentials for some states above or near
the corresponding meson thresholds. Such  components in addition to the
main one $QQQ$ could explain e.g. an exceptionally big splitting of the
flavor singlet states $\Lambda(1405)-\Lambda(1520)$, since the $\Lambda(1405)$
lies below the $\bar K N$ threshold and can be presented as  $\bar K N$ bound
system \cite{DAL}. Note, that in the case of the present approach this old
idea is completely natural and does not contradict a flavor-singlet  $QQQ$ 
nature of $\Lambda(1405)$, while it would be in conflict with naive constituent
quark model where no room for mesons in baryons. 
An admixture of such components
will be  important in order to understand strong decays of some excited states.
While technically inclusion of such  components in addition to the main
one $QQQ$ in a coupled-channel approach is rather difficult task, it should
be considered as one of the most important future directions.

What is an intuitive picture of the nucleon in the low-energy regime? 
The Goldstone bosons as well as vector meson fields 
couple to valence quarks. Thus the
nucleon consists mostly of 3 constituent quarks which are very big objects due 
to their meson clouds (see fig. 3). These constituent quarks are all the time 
in strong overlap inside the nucleon. That is why the short-range part of 
chiral interaction 
(which is represented by the contact term in the oversimplified 
representation ) is so crucially important inside baryons.
When constituent quarks are well separated and there is a phase space for 
meson propagation, the long-range Yukawa tails of 
meson-exchange interactions become very important. It is these parts
of meson exchange which produce the necessary long- and intermediate-range 
attraction in two-nucleon system.

\section{The Baryon-Baryon Interaction in a Chiral Constituent Quark Model}

If one ascribes a short-range central repulsion in the NN system to the 
central part of $\omega$ exchange between nucleons, 
then one should increase the 
$\omega N$ coupling constant by factor 3 as compared to its empirical 
value, as it is usually done in phenomenological one-boson-exchange NN 
potentials. Evidently the short-range repulsion in the NN system should
be connected with the nucleon structure in the low-energy regime 
and within the quark picture
should be related to fermi-nature of constituent quarks and to the
specific interactions between them.

So far, all studies of the short-range $NN$ interaction within the
constituent quark model were based on the one-gluon exchange interaction
between quarks. They explained the short-range repulsion in the $NN$ system
as due to the colour-magnetic part of OGE combined with quark interchanges
between 3Q clusters \cite{OKA}. It has been shown, however, that there
is practically no room for colour-magnetic interaction in light 
baryon spectroscopy and any appreciable amount of colour-magnetic interaction, 
in addition to chiral interaction, 
destroys the spectrum \cite{GPPVW}. This conclusion
is confirmed by recent lattice QCD calculations \cite{LIU}. If so, the
question arises which interquark interaction is responsible for the
short-range $NN$ repulsion. Below I show that the same short-range part
of chiral interaction (\ref{1.1}) 
which causes e.g. $N-\Delta$ splitting and produces good baryon spectra,
also induces a short-range repulsion in $NN$ system when the latter is
treated as 6Q system \cite{STPEPGL}.

At present one can use only a simple nonrelativistic $s^3$ ansatz for
the nucleon wave function when one applies the quark model to $NN$ interaction.
Thus one needs first an effective nonrelativistic parametrization
of chiral interaction interaction which  would provide correct
nucleon mass, $N-\Delta$ splitting and the nucleon stability with this ansatz. 
For that one
can use the nonrelativistic parametrization \cite{GPP}
which satisfies approximately  the conditions above.

In order to have a qualitative insight into the  $NN$ interaction
it is convenient to use an adiabatic Born-Oppenheimer approximation
for the internucleon potential:

\begin{equation} V_{NN}(R) = <H>_R - <H>_\infty, \label{6} \end{equation}

\noindent
where $R$ is a collective (generator) coordinate which is the separation
distance between the two wells (it should not be mixed with the relative
motion Jacobi coordinate), $<H>_R$ is the lowest
expectation value of the 6Q Hamiltonian at fixed $R$, and $<H>_\infty$
is a mass of two well-separated nucleons ($2m_N$) calculated with the same
Hamiltonian.

At the moment we are interested in what is the $NN$ interaction at zero
separation between nucleons. It has been proved by Harvey that when
$R\rightarrow 0$, then in both
$^3S_1$ and $^1S_0$ partial waves in the $NN$ system only two types
of orbital 6Q configurations survive \cite{HARVEY}: 
$|s^6 [6]_O>$ and  $|s^4p^2 [42]_O>$, where $[f]_O$ is Young diagram,
describing spatial permutational symmetry in 6Q system. There are
a few different flavor-spin symmetries, compatible with the spatial symmetries
above: $[6]_O[33]_{FS}$, $[42]_O[33]_{FS}$, $[42]_O[51]_{FS}$, 
$[42]_O[411]_{FS}$, $[42]_O[321]_{FS}$, and $[42]_O[2211]_{FS}$. Thus,
in order to evaluate the $NN$ interaction at zero separation between
nucleons  it is necessary to
diagonalize a $6Q$ Hamiltonian in the basis above and use 
the procedure (\ref{6}).

From the adiabatic Born-Oppenheimer approximation (\ref{6}) we find
that $V_{NN}(R=0)$ is highly repulsive in both $^3S_1$ and $^1S_0$
partial waves, with the core being of order 1 GeV. 
This repulsion implies a strong suppression of the $NN$ wave function 
in the nucleon overlap region.

Due to the specific flavor-spin symmetry of chiral interaction between quarks 
the configuration
$s^4p^2[42]_O[51]_{FS}$ becomes highly dominant among other possible 6Q 
configurations at zero separation between nucleons (however, the "energy"
of this configuration is much higher than the energy of two well-separated
nucleons, that is why there is a strong short-range repulsion in 
$NN$ system). The symmetry structure of this
dominant configuration induces "an additional" effective repulsion,
related to the "Pauli forbidden state" in this case, and the s-wave 
$NN$ relative motion wave function has a node at short range\cite{NST}.
The existence of a strong repulsion, related to the energy ballance, discussed
above, suggests, however, that the amplitude of the
oscillating $NN$ wave function at short range will be strongly suppressed. 

Thus, within the chiral constituent quark model one has all the necessary
ingredients to understand microscopically the $NN$ interaction. There
appears strong short-range repulsion from the same short-range part of 
chiral interaction
which also produces hyperfine splittings in baryon spectroscopy.
The long- and intermediate-range attraction in the $NN$ system is
automatically implied by the Yukawa part of pion-exchange and correlated
two-pion exchanges ($\sigma$-exchange) between quarks belonging to 
different nucleons. The necessary tensor and spin-orbit force comes 
from the pseudoscalar- and vector-exchange Yukawa tails.

What will be a short-range interaction in other $YN$ and $YY$ systems?
In the chiral limit there is no difference between all octet baryons:
$N, \Lambda, \Sigma, \Xi$. Thus if one explains a strong short-range
repulsion in $NN$ system as related mostly to the spontaneous breaking
of chiral symmetry, as above, then the same short-range repulsion
should persist in other $YN$ and $YY$ systems \cite{H}. Of course, due to the
explicit chiral symmetry breaking the strength of this repulsion should
be essentially different as compared to that one in $NN$ system. One
can naively expect that it will be weaker.

\bigskip
\noindent
{\bf Acknowledgement}

It is my pleasure to thank D.O.Riska, Z.Papp, W.Plessas, K.Varga, 
R. Wagenbrunn, Fl. Stancu, and S. Pepin, in collaboration with whom 
different results
discussed in this talk have been obtained. I also thank organizers of 
the school for the invitation to give this lecture.

\end{document}